\documentclass{mn2e}
\usepackage{amssymb}
\usepackage{graphicx}
\usepackage{amsmath}

\begin{document}

\title[PLC Relation for Early-Type Contact Binaries]
{Period-Luminosity-Colour Relation for Early-Type Contact Binaries}

\author[M. Pawlak]{Micha{\l} Pawlak \\
Warsaw University Observatory, Al. Ujazdowskie 4, 00-478 Warszawa, Poland}

\maketitle
\label{firstpage}


\begin{abstract}

This work describes the analysis of 64 early-type, massive contact or near-contact
eclipsing systems from the Large Magellanic Cloud discovered by the OGLE-III survey. It presents the determination of 
the period-luminosity-colour relation followed by these objects, that is different from the one previously known for late-type W UMa stars. 
The relation for massive stars has a significantly steeper dependence on the colour, which is related to a much higher bolometric 
correction, however it is shallower in the period term. 
This leads to the conclusion, that the relation for the total population of main sequence contact binaries is non-linear. 
When studied separately, genuinely-contact and near-contact systems follow two slightly different relations.
\end{abstract}

\begin{keywords}
stars: binary: eclipsing, stars: early-type, Magellanic Clouds
\end{keywords}

\section{Introduction}

Contact and close binary stars are known to form period-luminosity (PL) or period-luminosity-colour (PLC) relations due to 
the correlation between the radius of the orbit and the size of the components of the system.
Systems with an ellipsoidal red giant component filling its Roche lobe form a PL relation in the range of periods
from several to about 1000~days. It was discovered by Wood et al. (1999) and studied by Rucinski and Maceroni (2001),
Soszy{\'n}ski et al. (2004, 2007) and most recently by Pawlak et al. (2014).

Another example is the relation formed by main sequence contact binary systems.
The PLC relation for low-mass W UMa type stars have been studied in detail by Rucinski (1994, 2002, 2004),
who used nearby stars with known parallaxes in order to calibrate the relation.
While the scatter of the absolute magnitude derived from the PLC relation for W UMa stars is
larger than for the pulsating stars, it can be very useful as a distance ruler 
because contact binaries are common and easy to detect with the time-series photometry. 
This method was used, for example, to verify cluster membership of W UMa type stars 
(Maciejewski et al. 2008, Kopacki et al. 2008, Li et al. 2010, Hu et al. 2011, Joshi et al. 2012).

While the PLC relation for late-type W UMa stars can be analysed based on the nearby objects, the situation
is different for more massive stars. These are less abundant, therefore we do not have a
statistically significant sample of such objects with known parallaxes. However, a large number of contact
binaries were discovered during the third phase of the Optical Gravitational Lensing Experiment 
(OGLE-III, Udalski 2003) project in the Large Magellanic Cloud (Graczyk et al. 2011), Small Magellanic 
Cloud (Pawlak et al. 2013) and Galactic disc (Pietrukowicz et al. 2013). The LMC sample is especially 
useful because of the well-known distance to this galaxy (Pietrzy{\'n}ski et al. 2013).

In this work, the analysis of early-type contact eclipsing binaries in the LMC is performed.
The PLC relation they follow is found and compared with the results for the low-mass systems (Rucinski 2004).
The structure of the paper is as follow: Section~2 gives the details on the analysed data, Section~3
presents the PLC relation obtained for the studied sample, Section~4 contains the discussion
and Section~5 summarizes the results.

\section{The Sample}

The analysis is based on the sample of contact or very close systems from the OGLE-III catalogue of eclipsing binaries
in the LMC (Graczyk et al. 2011). Out of the entire collection of light curves classified as contact, 
a subset of 64 systems having well-covered light curves, with low photometric noise, 
and a shape typical for a contact or near-contact binaries, was extracted. 
For such objects the transition from the minimum to the maximum of magnitude
is smooth, and it is impossible to distinguish the eclipse from out-of-eclipse phase,
even with precise photometry (Fig.~1).

\begin{figure}
\label{fig:1}
\includegraphics[angle=270,width=41mm]{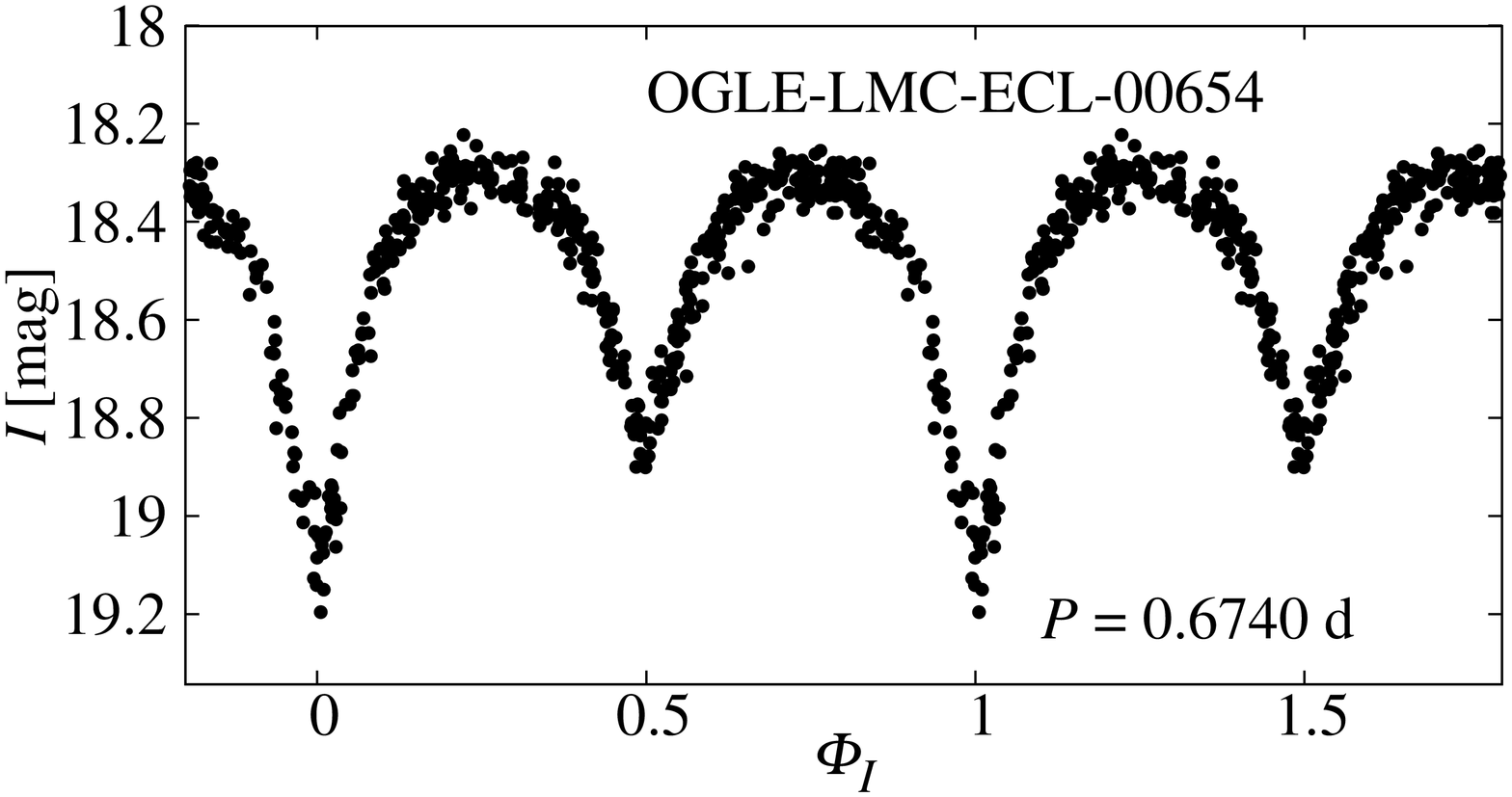} \includegraphics[angle=270,width=41mm]{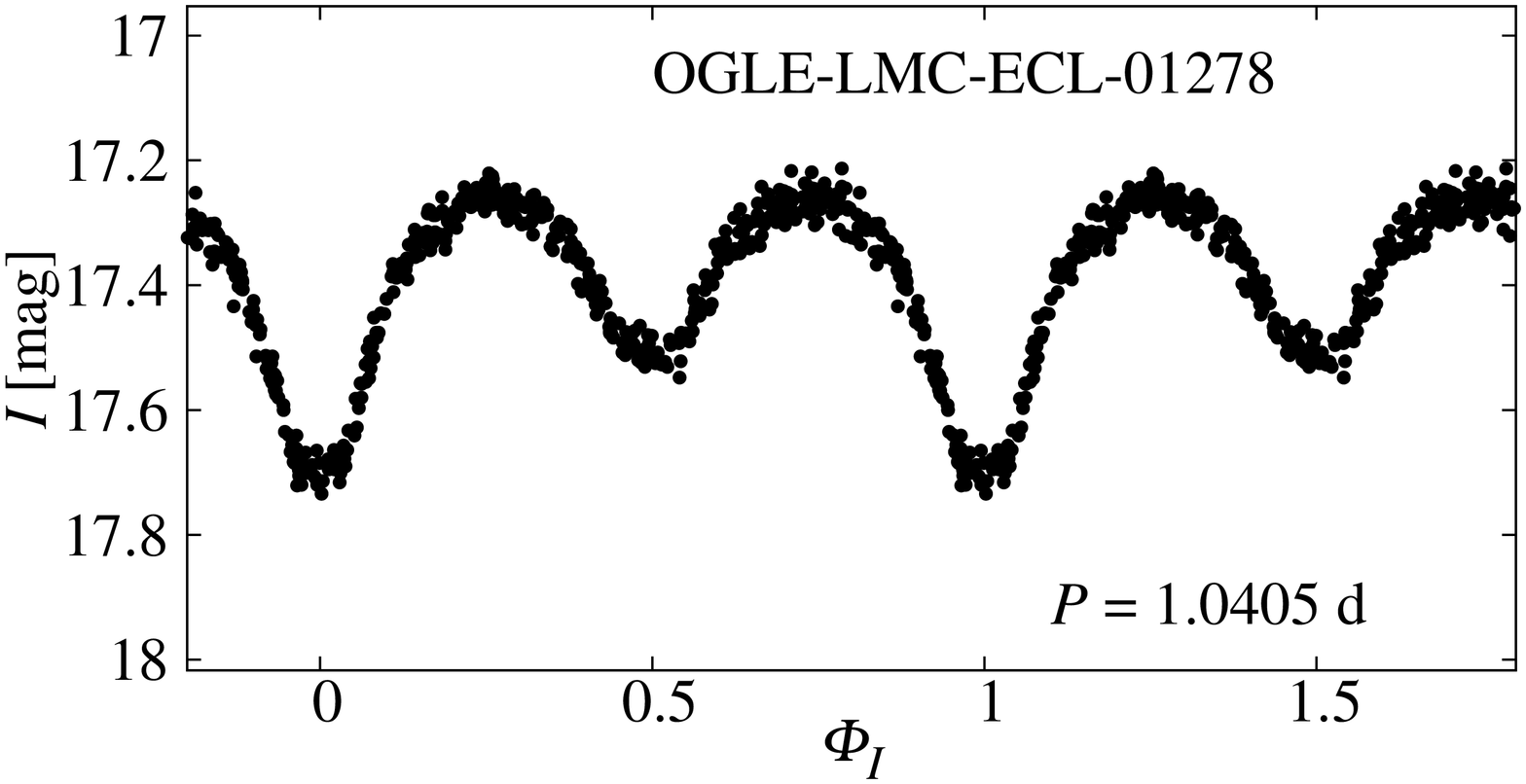} \\
\includegraphics[angle=270,width=41mm]{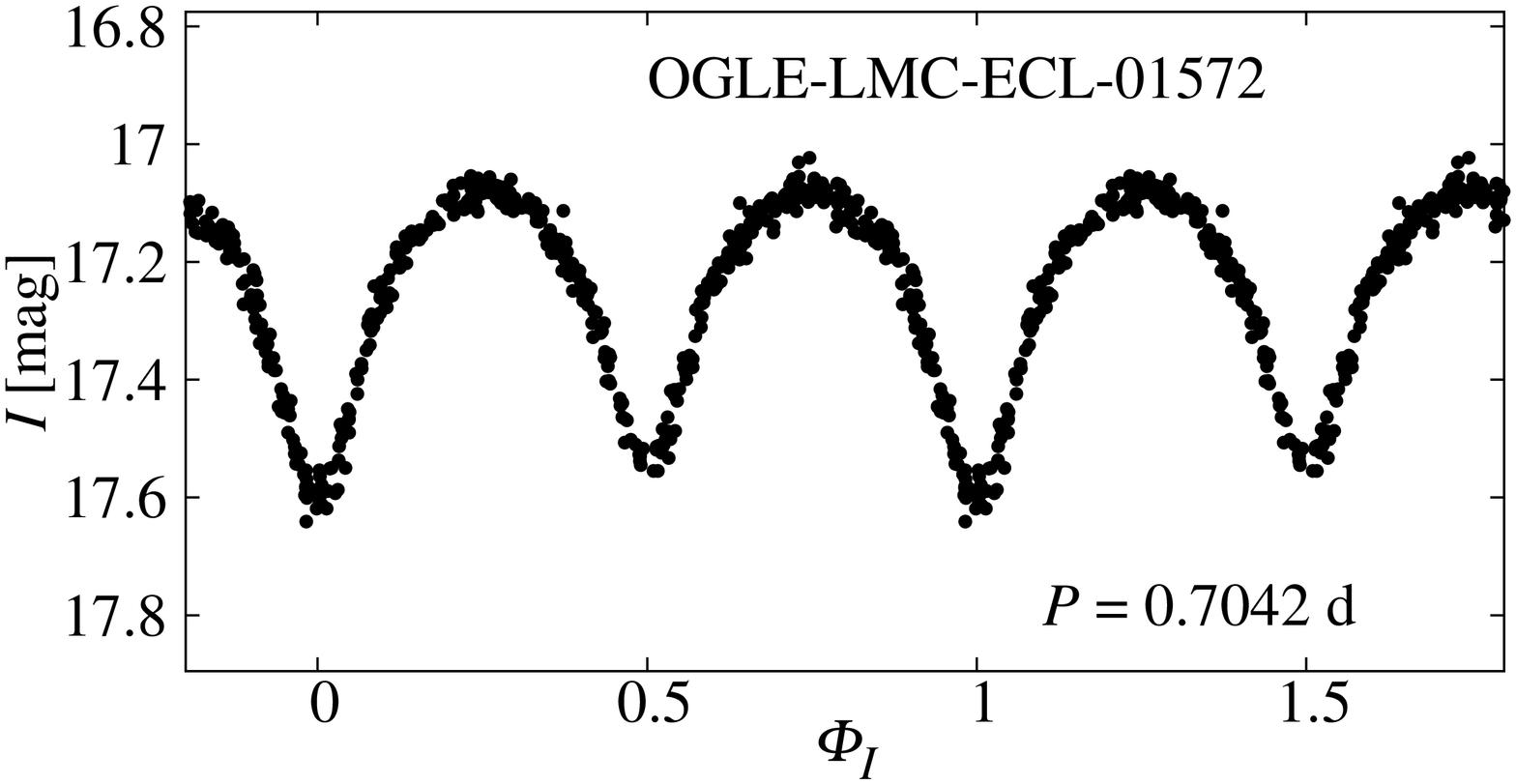} \includegraphics[angle=270,width=41mm]{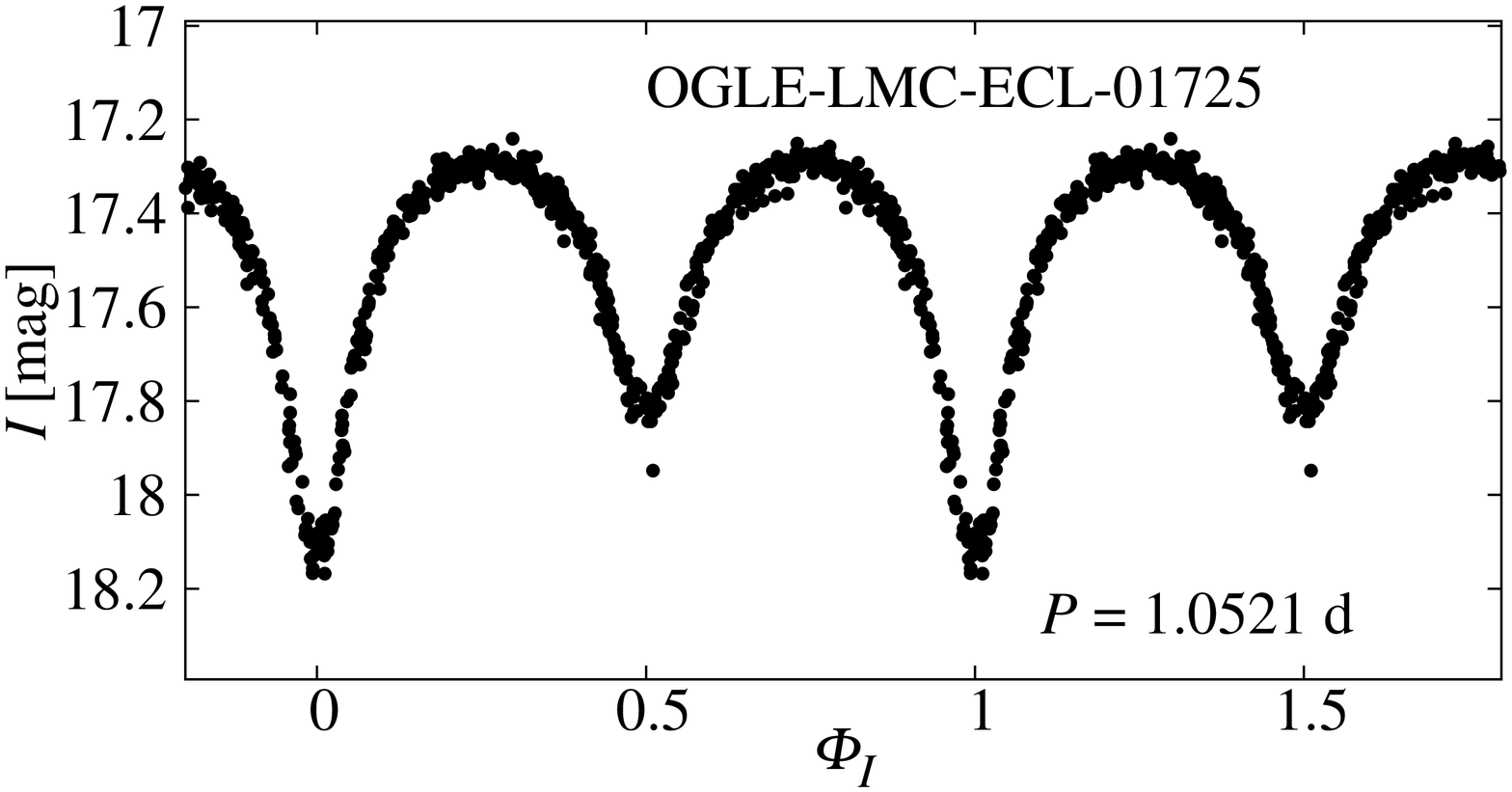} \\
\includegraphics[angle=270,width=41mm]{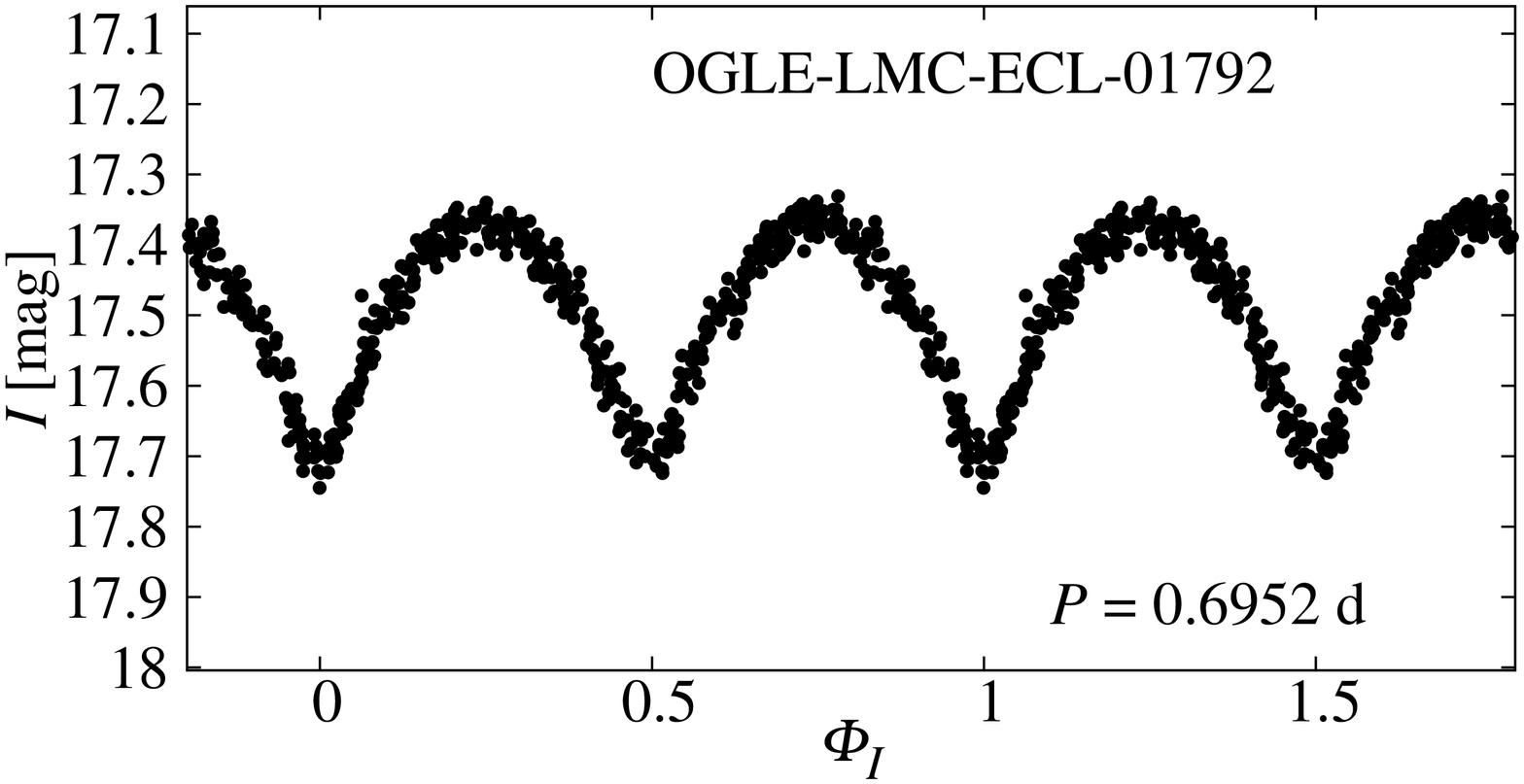} \includegraphics[angle=270,width=41mm]{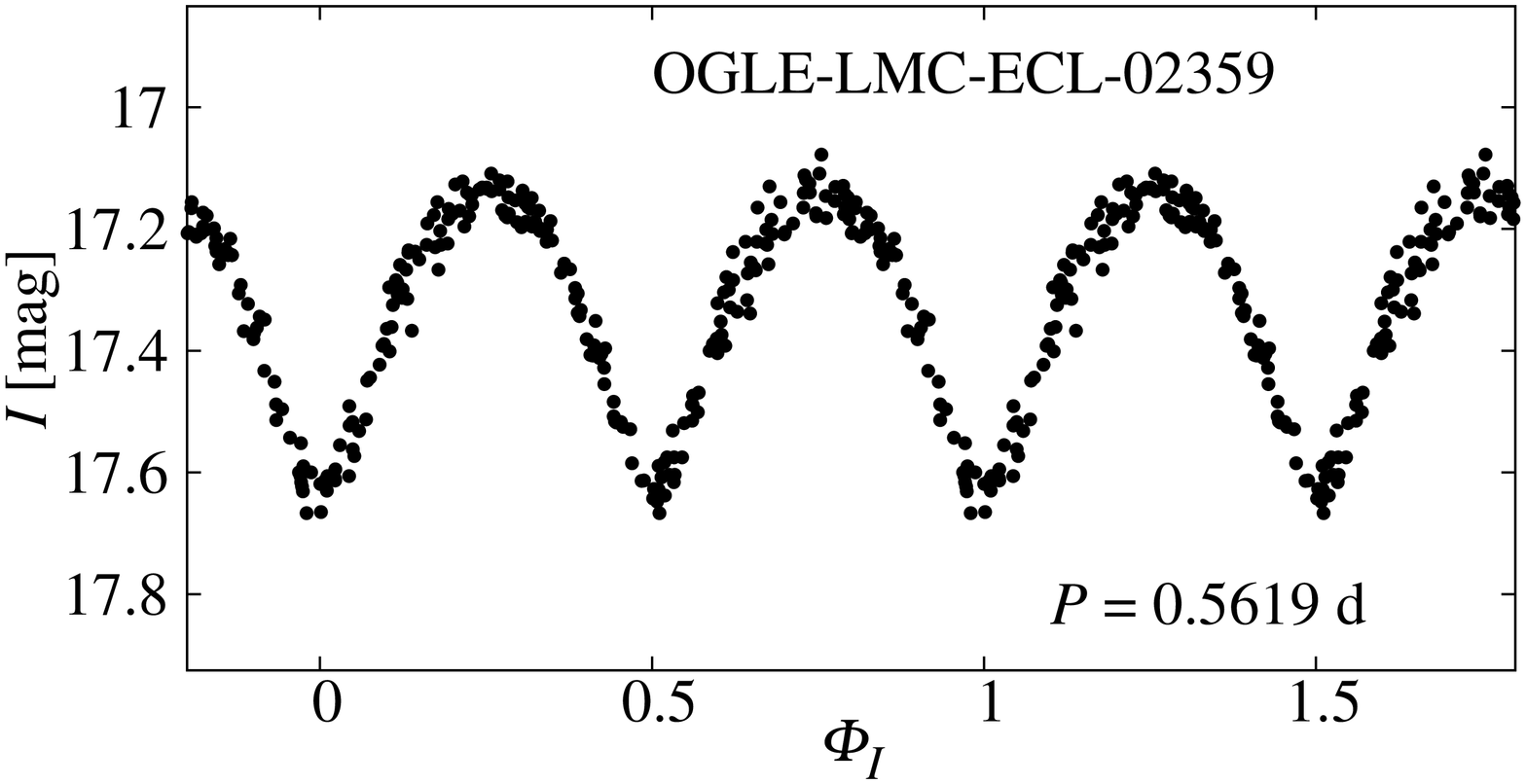} \\
\includegraphics[angle=270,width=41mm]{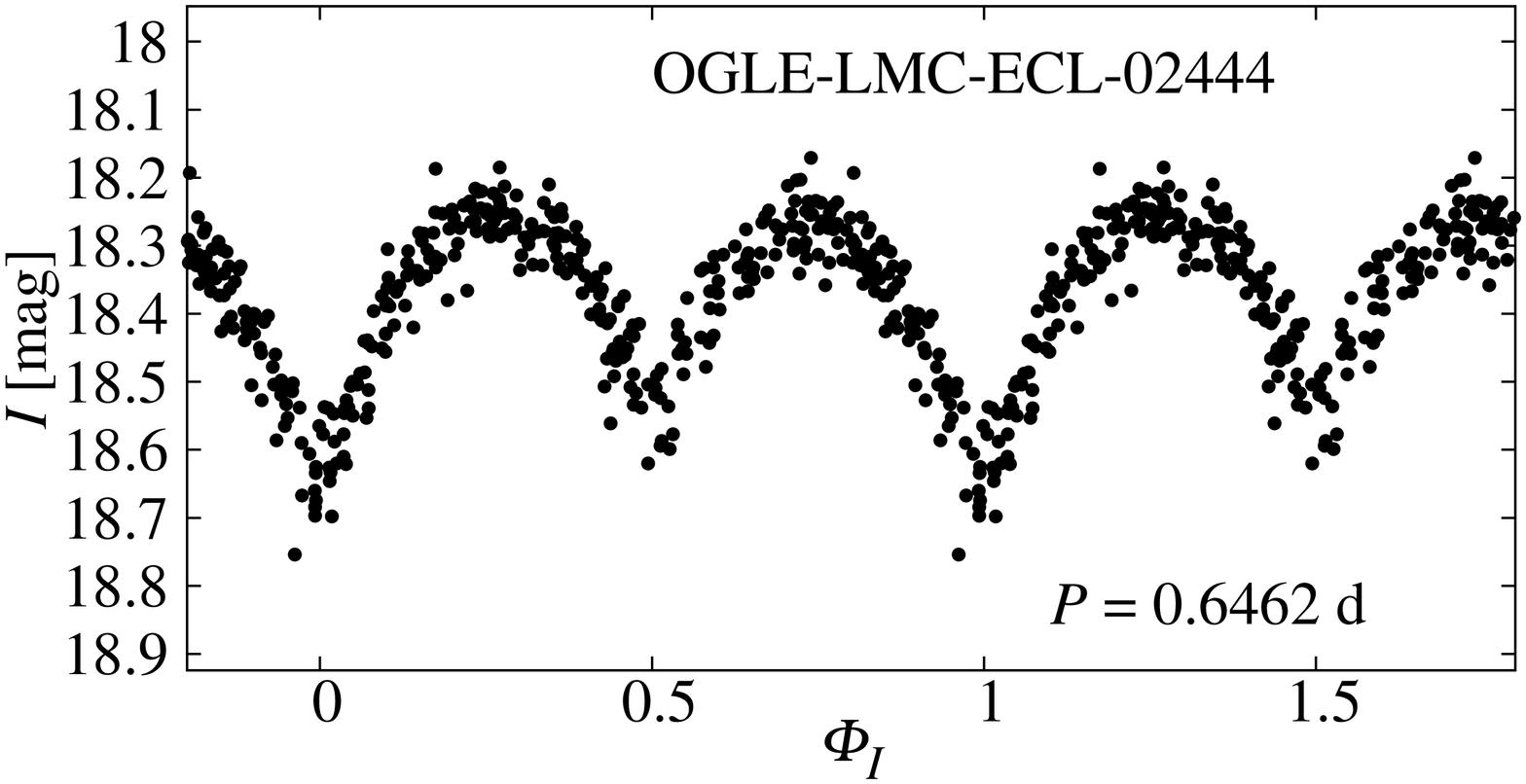} \includegraphics[angle=270,width=41mm]{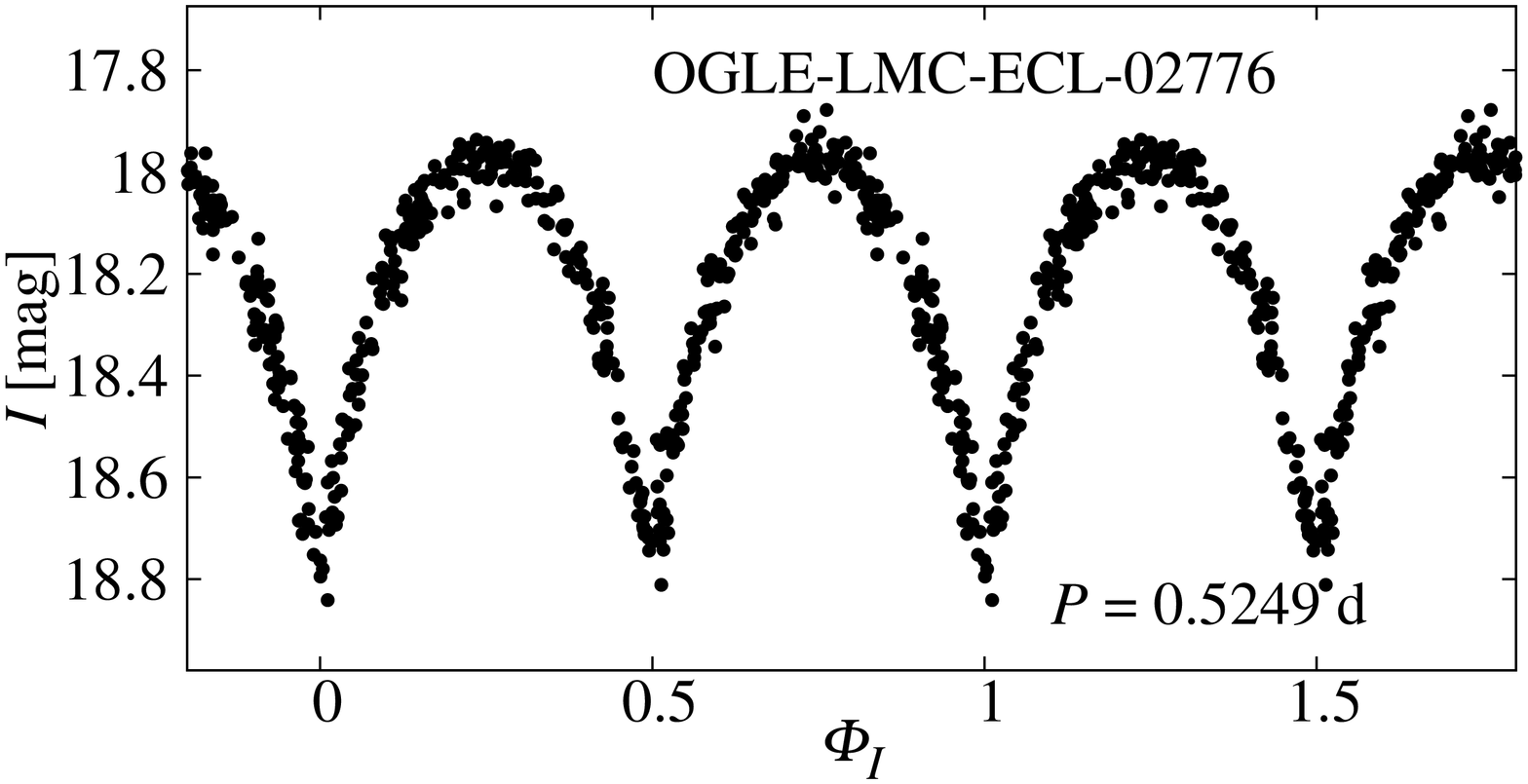} \\
\includegraphics[angle=270,width=41mm]{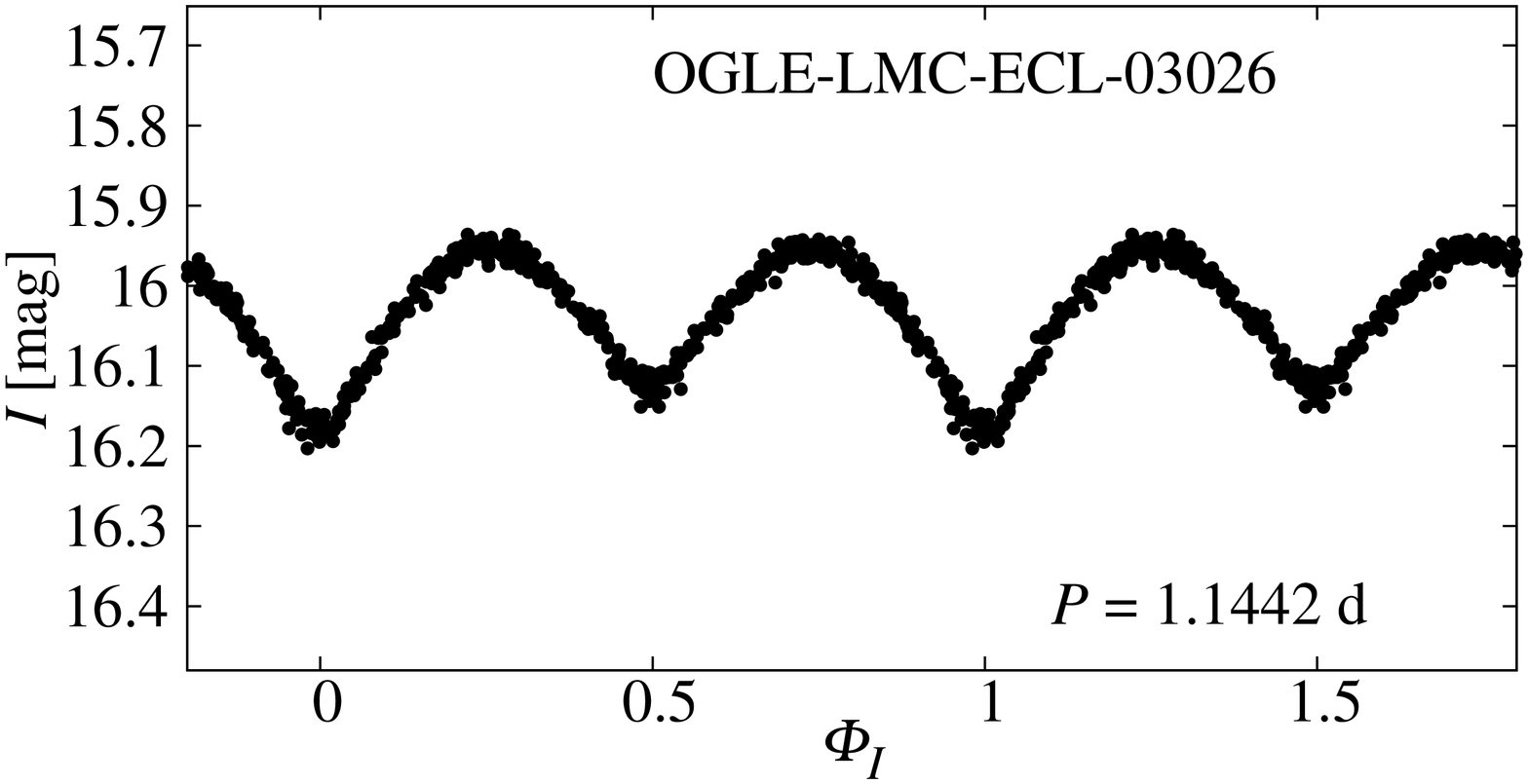} \includegraphics[angle=270,width=41mm]{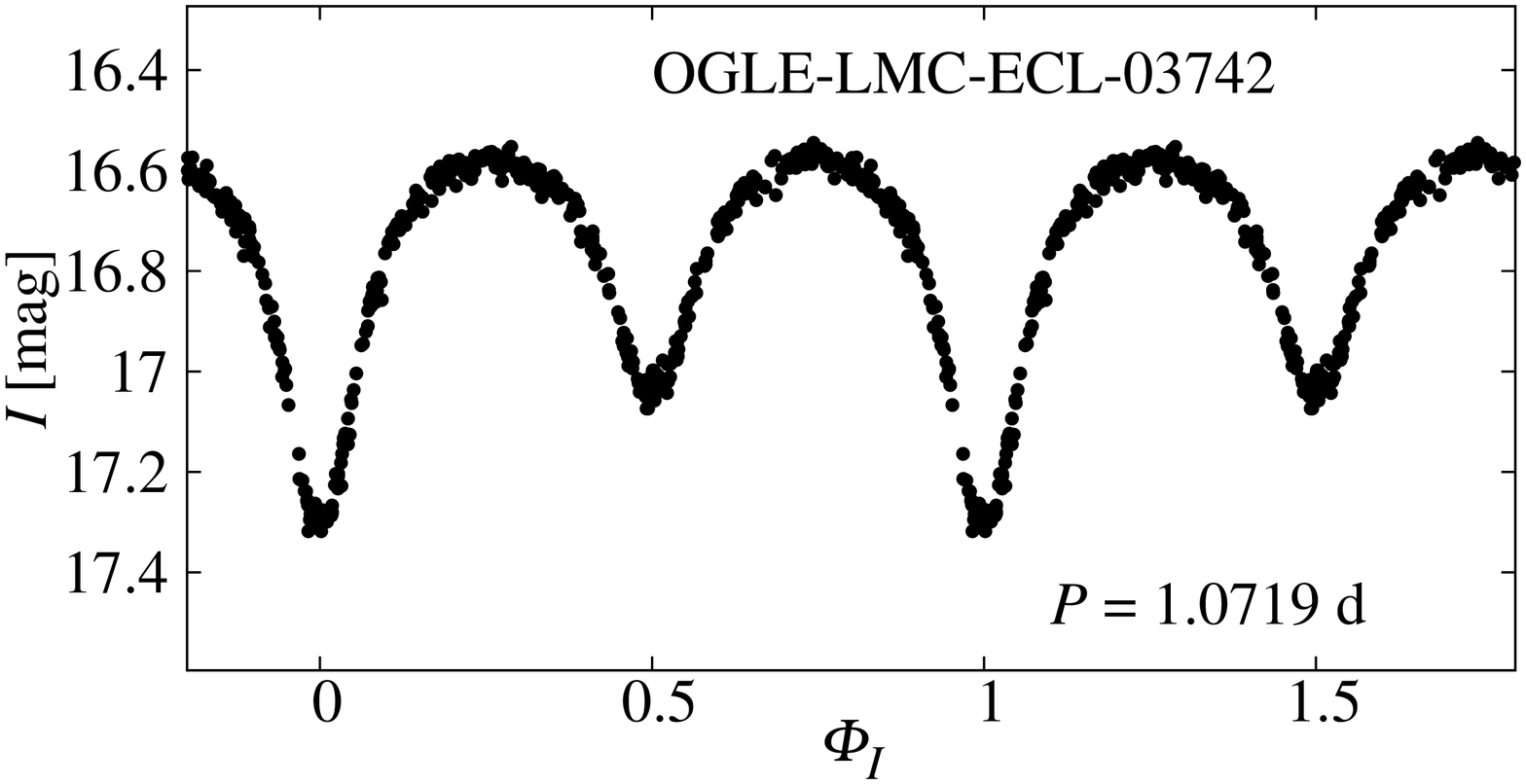} 
 
\caption{Example light curves of selected contact or close eclipsing binaries from the OGLE-III catalogue of
eclipsing binaries in the LMC (Graczyk et al. 2011).}
\end{figure}

Fig.~2 shows the positions of the selected objects in the colour-magnitude diagram. All of selected objects
are bright and located in the LMC main sequence region or close to it. Therefore, these are most likely massive, 
early-type stars belonging to the LMC, and not foreground objects.

\begin{figure}
\label{fig:2}
\centerline{
\includegraphics[angle=270,width=90mm]{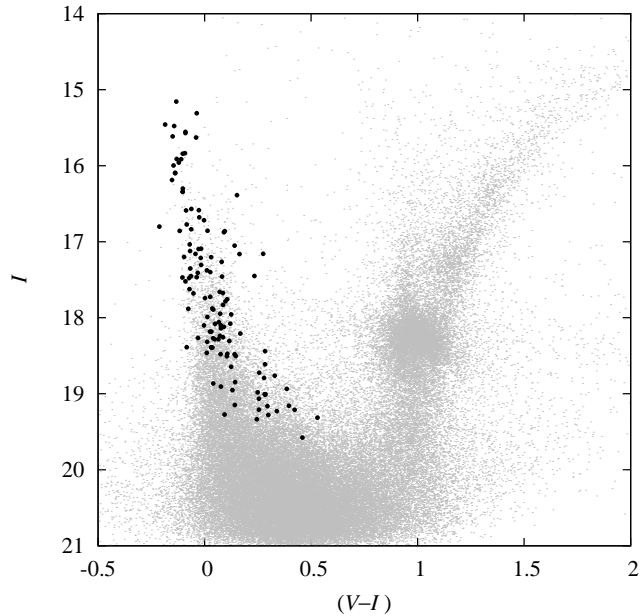}}
\caption{Colour-magnitude diagram for the OGLE-III LMC data, with marked contact binaries from the studied sample (black points).
Selected objects clearly lie on the LMC main sequence or close to it. The field stars (\textbf{grey} dots) come from the OGLE-III 
photometric maps from one of the central LMC fields (Udalski et al. 2008).}
\end{figure}

All of these objects lie in the area of the sky corresponding to the bar of the LMC, where the bulk of
the OGLE-III coverage of the LMC was concentrated. 
Moreover, these are early-type stars belonging to the young population. Therefore, we expect them to be 
located in the centre of the LMC rather than in its halo. 
The geometric extent of the young population in the LMC can be limited by looking at
the scatter of the PL relation for Classical Cepheids, which is equal to 0.07 mag (Soszy{\'n}ski et al. 2008). 
This value was obtained using Wesenheit index, therefore
it is reddening-independent and corresponds only to geometric effects and internal scatter of the relation.  
The low geometric dispersion allows us to adopt for each of the systems
the distance modulus ($DM$) of $18.49\pm0.05$~mag, obtained by Pietrzy{\'n}ski et al. (2013).

To calculate the absolute magnitude of each of the systems, the correction for the interstellar extinction
is needed. The estimation of total extinction in the $V$-band ($A_V$) is done based on
the LMC reddening maps (Haschke et al. 2011), using the transformation from $E(V-I)$ to $A_V$ from Schlegel et al. 1998.
To provide a better $A_V$ estimation, the mean Red Clump (RC) colour was measured in the OGLE-IV
fields located in the outskirts of the LMC (Udalski et al. 2015). The obtained value of ${(V-I)}^{RC} = 0.91$~mag minus the 
mean foreground reddening $E(V-I) =0.04$~mag from Schlegel et al. (1998) results in the mean reddening-free colour of the RC
${(V-I)_0}^{RC} = 0.87$~mag. This value is adopted instead of the ${(V-I)_0}^{RC} = 0.92$~mag colour used by Haschke et al. (2011),
therefore the $E(V-I)$ obtained from the reddening maps is increased by the value of 0.05~mag, resulting from the difference
in the RC colour used.
The absolute $V$-band magnitude ($M_V$), in the maximum of the light curve for all of the studied systems, 
is calculated with the Eq.~\eqref{eq:1}.

\begin{equation}
\label{eq:1}
M_V = V - A_V -DM 
\end{equation}
 
The uncertainty in $M_V$ is determined by the uncertainty of $A_V$, which in the studied sample is on average 
0.19~mag. This is much higher than visual magnitude uncertainty or geometric scatter.

\section{Period-Luminosity-Colour Relation}

The PLC relation for late-type (LT) W UMa type variables (Rucinski 2004) is quoted in the Eq.~\eqref{eq:2}.

\begin{equation}
\label{eq:2}
M_{V,{\rm LT}} = -4.43\,\log P + 3.63\,(V-I)_0 -0.31
\end{equation}

This form of the PLC relation was already proposed by Rucinski (1994), as a simplification of a theoretically 
derived formula, including also the bolometric correction (BC) and the mass ratio ($q$) dependent terms, which
were neglected in the final calibration. 

The BC is very small for low-mass main-sequence stars, therefore it is 
not surprising that the PLC relation works well without explicitly including it. The situation is different
for massive objects, for which the BC becomes significant. However, the BC is a function of temperature and 
therefore the colour as its indicator. Moreover, Nieva (2013) shows that the dependency between BC and temperature 
is linear for hot stars. Therefore, one can expect that the PLC relation will work properly in this case without
explicit BC term, as this dependence is included in the colour term.  

Another question is whether the $q$-dependent term in the relation is explicitly necessary.
The precise determination of the $q$ value is difficult based on photometric data only. While $q$ can be estimated 
using a light curve synthesis code, such a solution requires the assumption of a given model (contact, semi-detached or detached),
therefore it is strongly model dependent and can introduce additional errors to the final calibration. Reliable determination 
of $q$ requires spectroscopic observations, but these are hard-to-obtain for faint binary systems.
Moreover, W UMa stars tend to have similar, low $q$, therefore this factor is not expected to contribute significantly 
to the relation. Because of that Rucinski (1994) decided to neglect the $q$-dependent factor.

The situation changes a bit for massive systems, which have $q$ spanning a wider range.
However, there still remains the problem of reliable $q$ determination. It cannot be done accurately based
on photometric data only, and requires spectroscopic observations, which are unavailable for the studied sample.
Therefore, this factor is also neglected in the calibration performed in this work. Further study, including 
spectroscopic analysis may allow for more precise calibration in this case.

Taking the above issues into account and
assuming the same form of the relation (Eq.~\eqref{eq:2}) for early-type (ET) contact systems, the linear function of log$P$ and
dereddened colour $(V-I)_0$ is fitted, with the least square method, to the whole studied sample. 
The best fit obtained is given by Eq.~\eqref{eq:3}.

\begin{align}
\label{eq:3}
M_{V,{\rm ET}} = &-2.97( \pm0.51)\,\log P \nonumber \\
                 &+8.27( \pm0.73)\,(V-I)_0 \nonumber \\ 
                 &-0.59( \pm0.15)
\end{align}

The distribution of the sample in the PLC space, together with the relation fitted, is presented in Fig.~3.
The systems clearly group in the plane given by the relation, however, its scatter is relatively large. The $RMS$ for 
the obtained fit is 0.52 mag, which is two times larger than in Rucinski (2004).

For further study,
the sample is visually divided into two groups: genuinely-contact systems, where the two components are in thermal equilibrium (light curves with
two equal minima), and near-contact systems, which are close, but still detached or semi-detached and therefore not in a thermal contact 
(light curves with minima of different depth). 
As it can be seen in Fig.~4, near-contact systems tend to lie below the PLC relation fitted for the entire sample.
This is consistent with a similar result obtained by Rucinski and Duerbeck (1997) for low-mass W UMa type stars.

Projections of the obtained PLC relation onto the ${\log}P$~$-$~$M_V$ and $(V-I)_0$~$-$~$M_V$ planes 
are shown in Fig.~5. It can be seen that the PL relation itself has a large scatter and does not allow 
for precise estimation of $M_V$ without adding the colour term, due to the strong correlation between colour 
and absolute magnitude resulting from higher BC. 
It also explains a relatively high uncertainty of the ${\log}P$ term in the PLC relation.     

The tight correlation of $M_V$ with $(V-I)_0$ and rather poor with ${\log}P$ arise a question if two 
independent variables are really needed to calibrate the relation. 
To test it a linear function of $(V-I)_0$ only is fitted to the data. The $RMS$ of such fit is 0.65~mag,  
which worse than the one obtained for a two parameter fit ($RMS = 0.52$~mag).
The decrease of scatter suggests that using the PLC relation is justified. However, it should be tested,
if this decrease is significant from the statistical point of view, as it can be related to the reduced 
number of degrees of freedom, caused by adding another parameter of the fit. 

To verify this, the $F$-test is performed for the residuals of the two fits. The value of the test statistic
for the two variances is $F = 1.52$ making it almost equal to the critical value for 62 and 61 degrees of freedom which is $F_{62,61} = 1.53$,
for the significance level $p = 0.05$. While the result of the test is a borderline case, it rather suggests to reject the hypothesis that the two variances 
are equal, what leads to the conclusion that the full PLC relation gives a better fit.

For further verification, the analysis of the residuals of the one-variable fit is done. If the period dependency was
absent, the residuals of colour-magnitude relation should be flat, without a noticeable correlation with the period.
However, such correlation is clearly visible in Fig.~6, where residual of the colour-magnitude fit are plotted
as a function of ${\log}P$. This strongly indicates, that the period dependency is significant in the proper calibration of the relation, 
even though it may at first look blurred by much stronger colour dependency.

Finally, the presence of the period therm is also consistent with the physical interpretation of the problem. 
All above leads to the conclusion, that two variables are indeed necessary.
Analogous question has also been discussed by Rucinski (2006) for W UMa stars where a similar conclusion was reached.

\begin{figure}
\label{fig:3}
\centerline{
\includegraphics[angle=270,width=86mm]{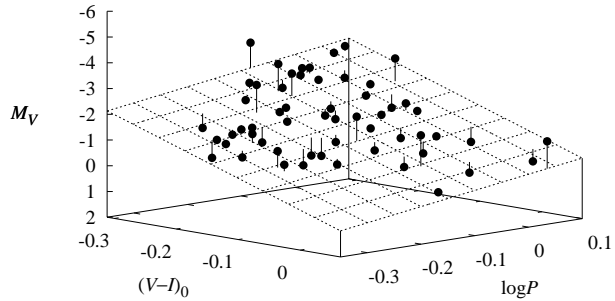}}
\caption{PLC relation obtained for the sample of 64 contact or near-contact, early-type binary systems from the LMC.}
\end{figure}

\begin{figure}
\label{fig:4}
\centerline{
\includegraphics[angle=270,width=86mm]{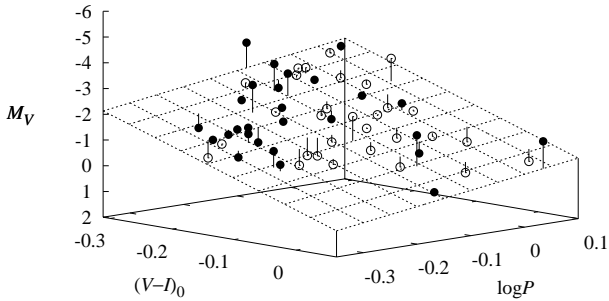}}
\caption{PLC relation for the sample divided into genuinely-contact (filled circles) and near-contact (open circles) systems. 
Near-contact binaries lie mostly bellow the relation obtained for the entire sample.}
\end{figure}

\begin{figure}
\label{fig:4}
\centerline{
\includegraphics[angle=270,width=86mm]{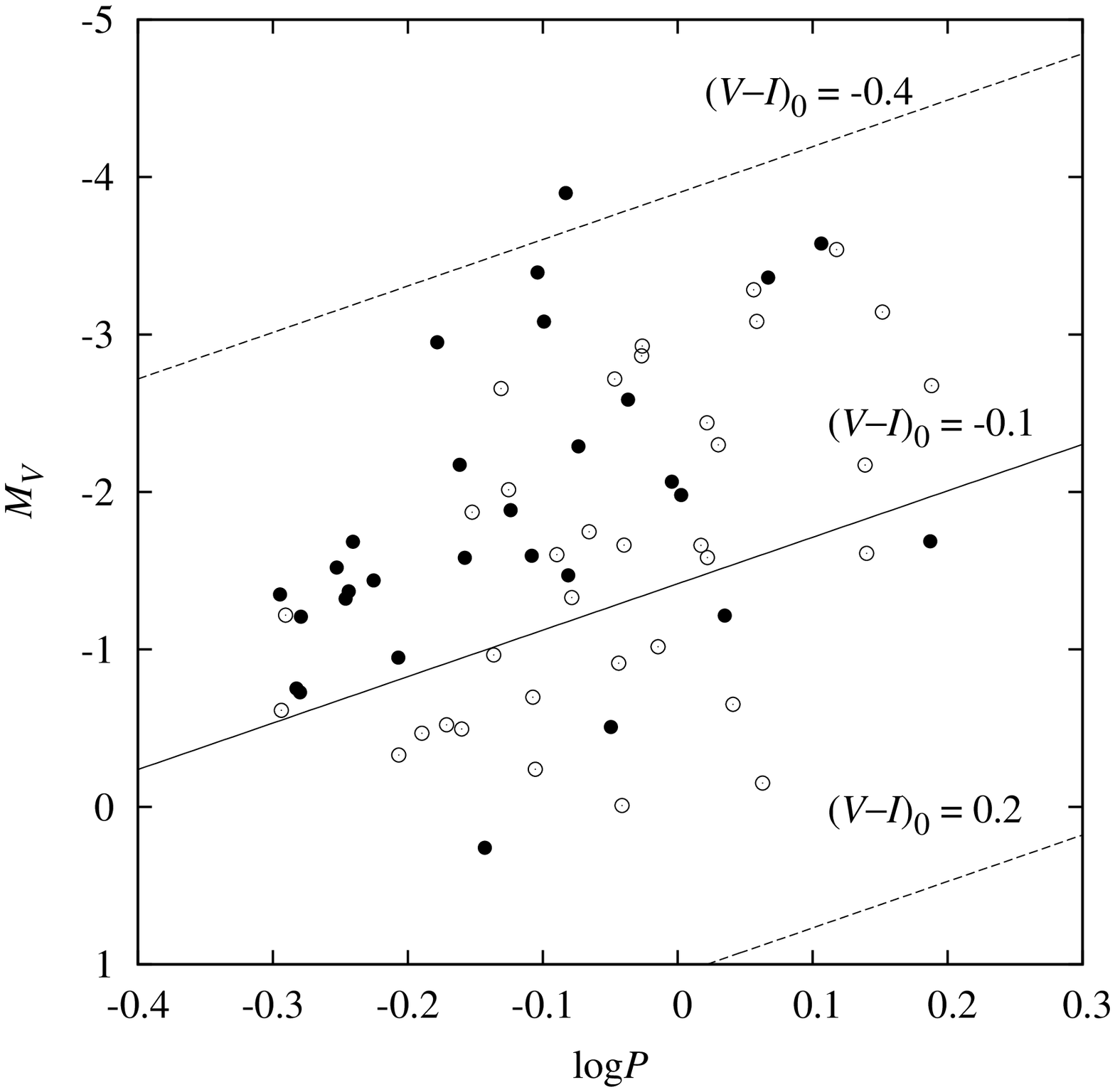}}
\centerline{
\includegraphics[angle=270,width=90mm]{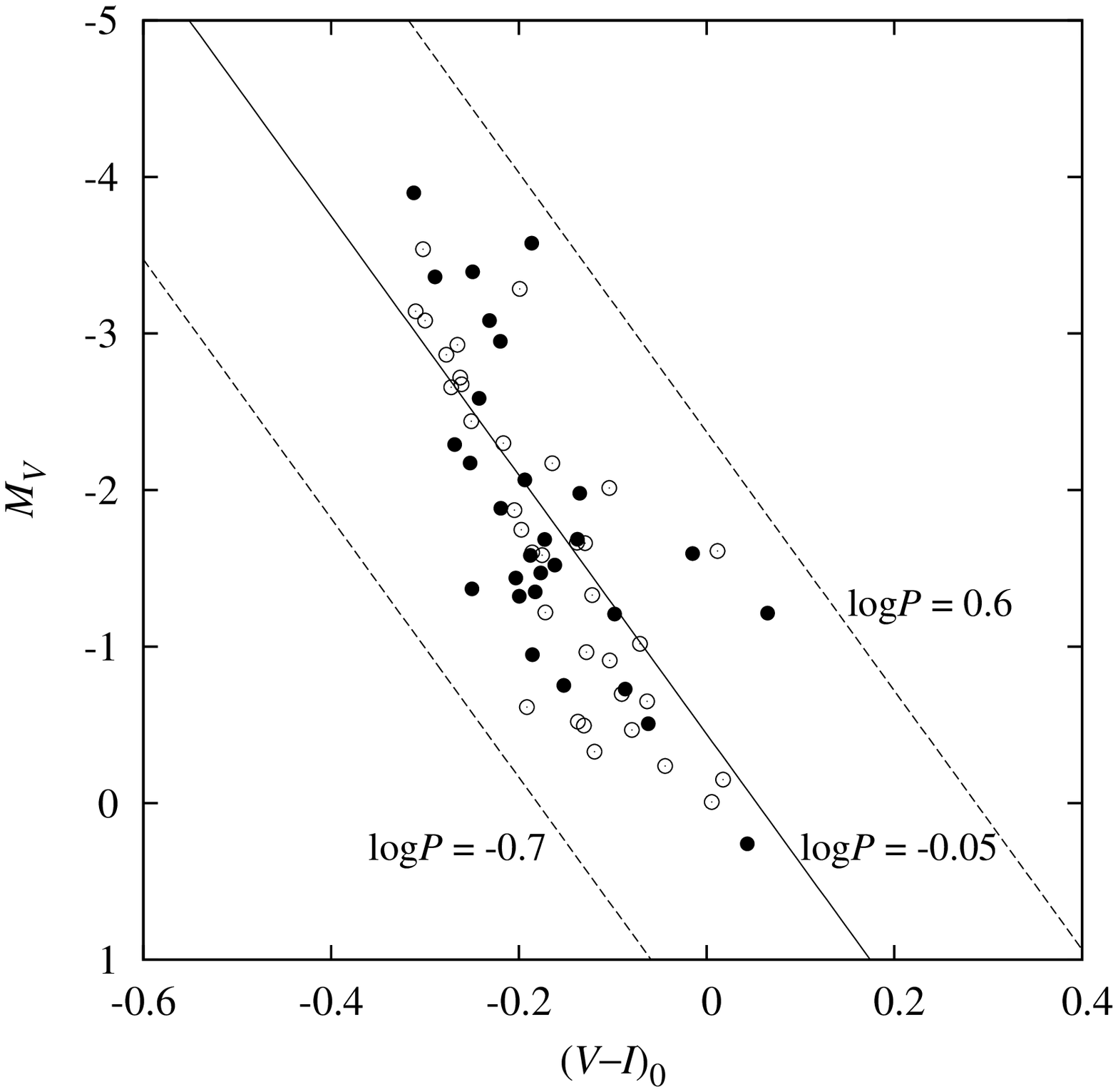}}
\caption{Projections of the PLC relation (see Fig.~3) into ${\log}P$~-~$M_V$ (upper panel) and $(V-I)_0$~-~$M_V$ plane (lower panel).
Projections of the PLC plane along three different lines are marked on each of the panels. Near-contact systems (open circles) 
tend to lie bellow genuienely-contact systems (filled circles)  in the upper panel (similara as in Fig. 4), 
while no significant separation is visible in the lower panel.}
\end{figure}

\begin{figure}
\label{fig:5}
\centerline{
\includegraphics[angle=270,width=86mm]{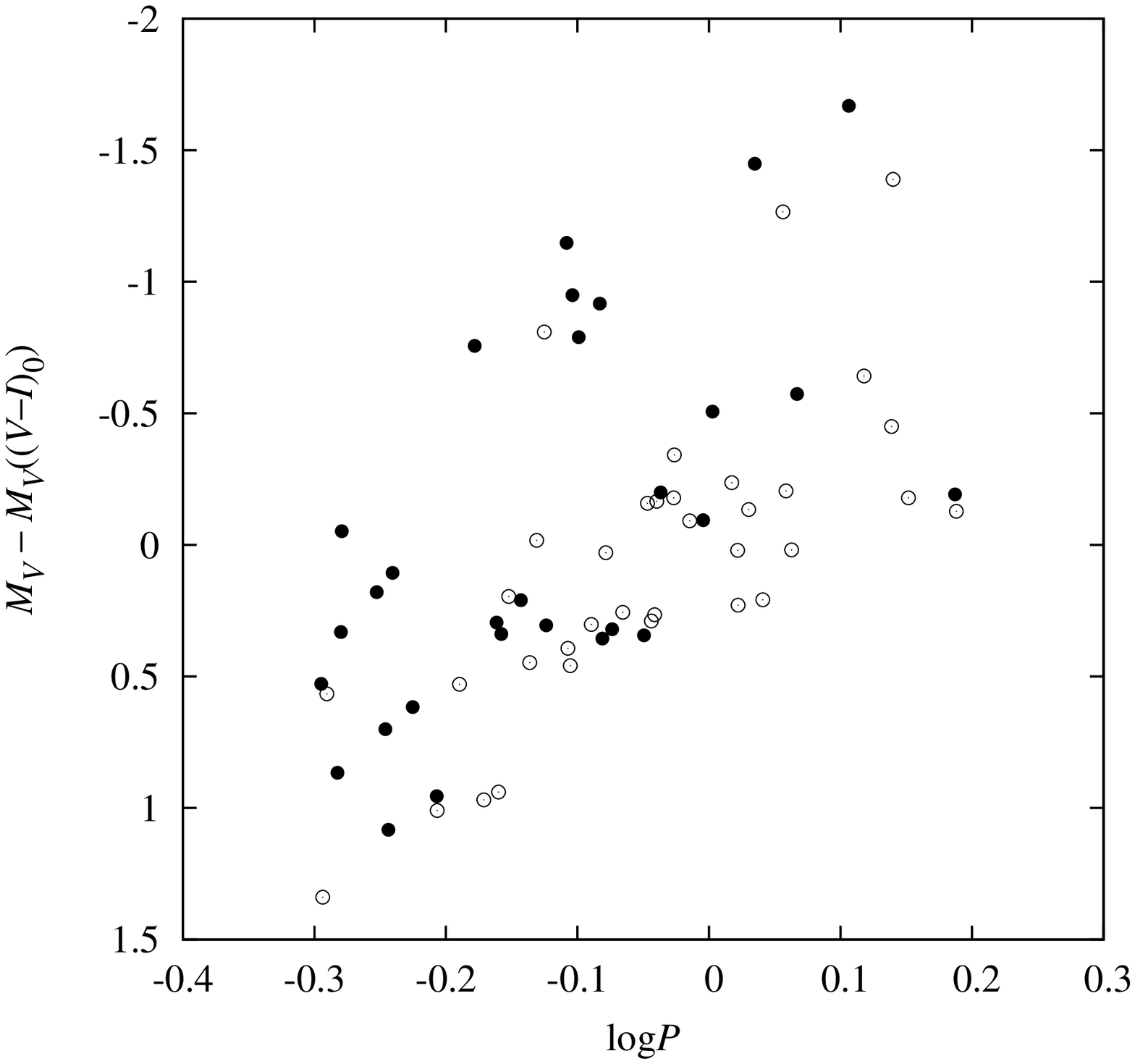}}
\caption{Residuals of the one variable fit as a function of ${\log}P$. A clear correlation is visible, showing that the relation using $(V-I)_0$ 
dependency only is not sufficient and that the period term is needed. One may suspect a division into the two types dependening on 
the type of the light curve, as expected for systematically differening sizes of genuinely-contact (filled circes) and near-contact (open circles) binaries.}
\end{figure}

In the next step, the PLC relation is fitted separately for each of the two previously defined groups (Fig.~7). The result is given by Eq.~\eqref{eq:4} 
for genuinely-contact (GC) systems and by Eq.~\eqref{eq:5} for near-contact (NC) systems.

\begin{figure}
\label{fig:7}
\centerline{
\includegraphics[angle=270,width=86mm]{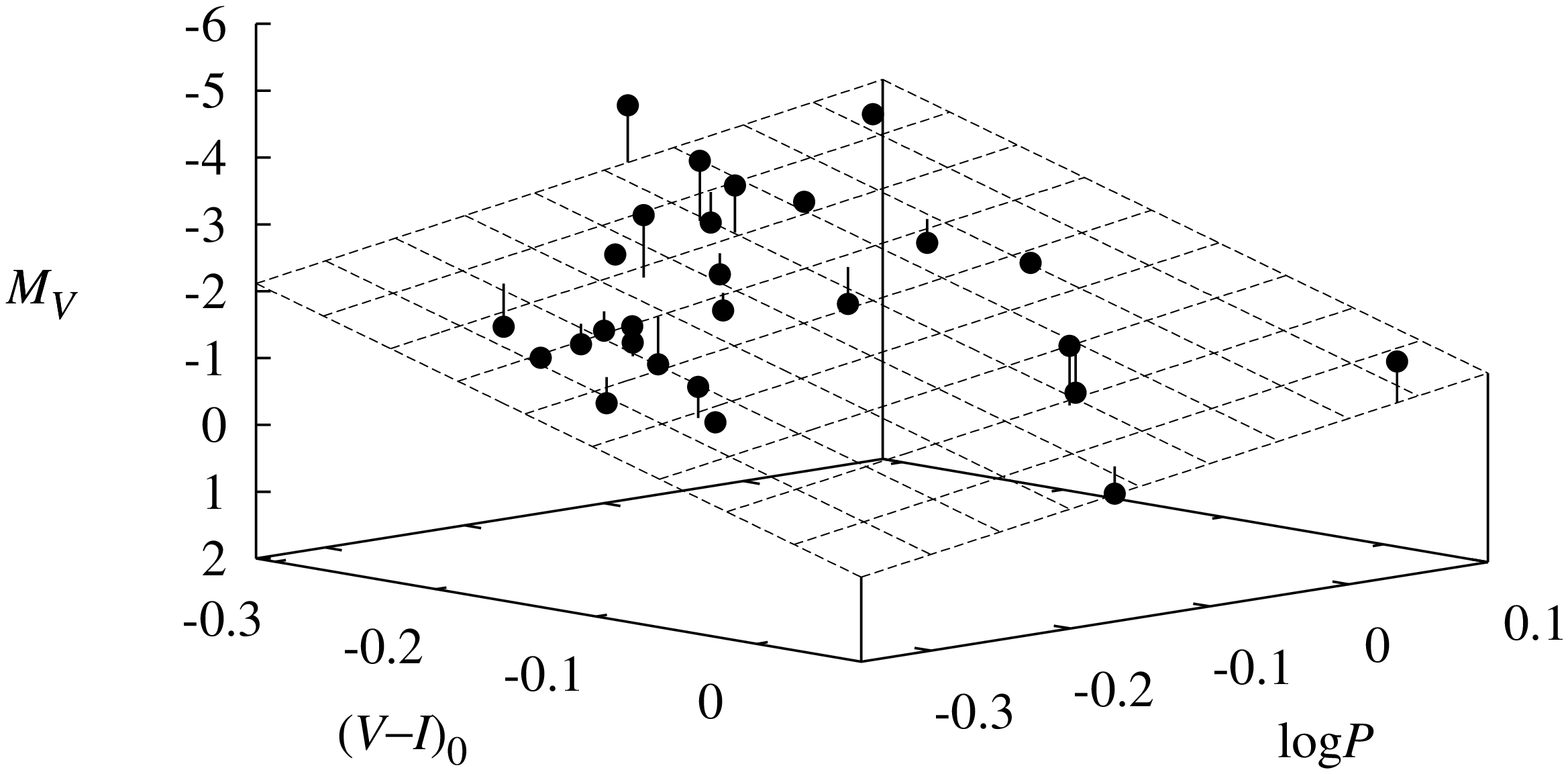}}
\centerline{
\includegraphics[angle=270,width=86mm]{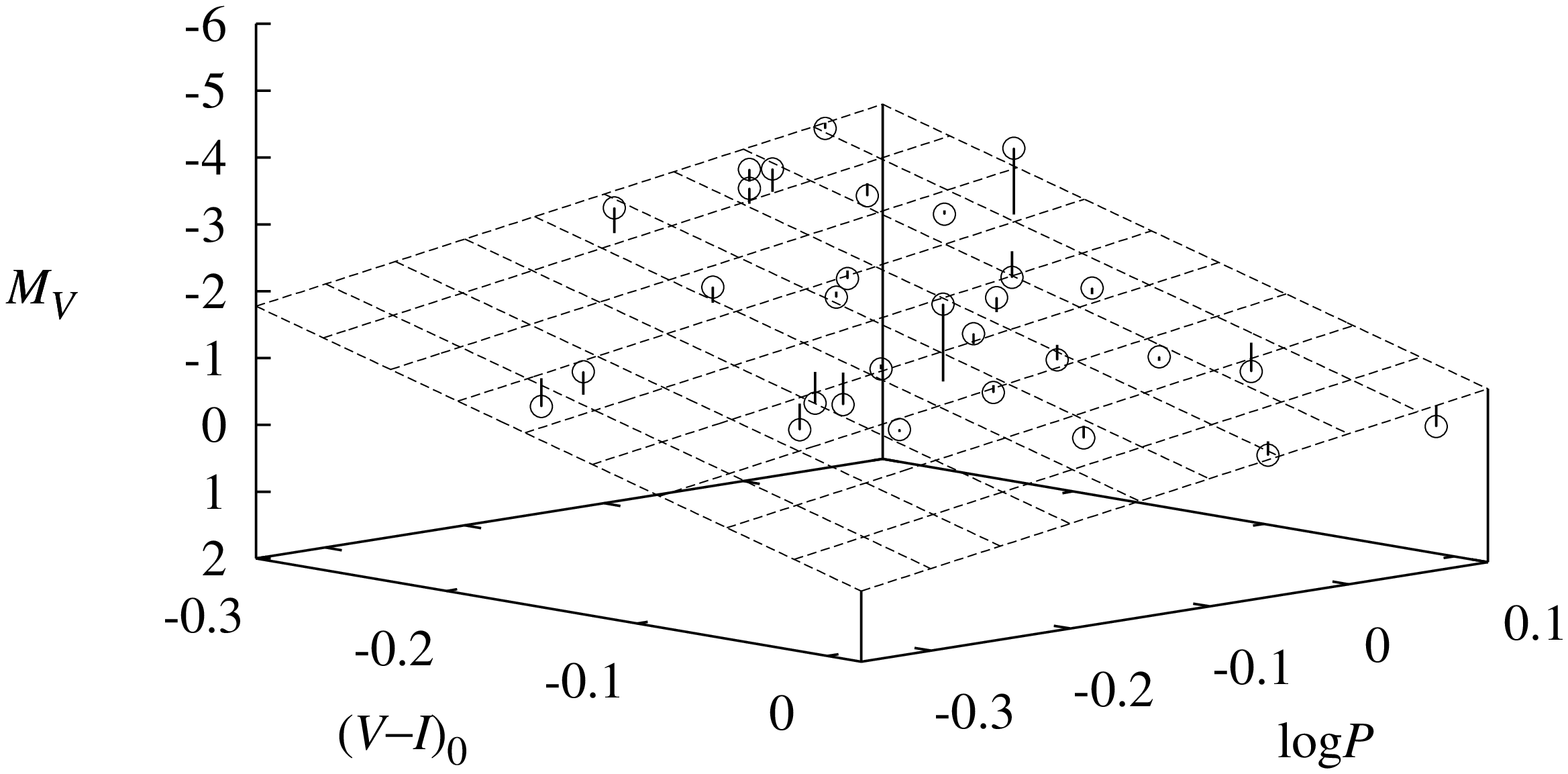}}
\caption{PLC relation fitted separately for genuinely-contact (upper panel) and near-contact (lower panel) systems.}
\end{figure}

\begin{align}
\label{eq:4}
M_{V,{\rm GC}} = &-3.47 (\pm0.87)\, \log P \nonumber \\
                 &+7.57 (\pm1.21)\,(V-I)_0 \nonumber \\
                 &-0.97 (\pm0.26)
\end{align}

\begin{align}
\label{eq:5}
M_{V,{\rm NC}} = &-3.41 (\pm0.60)\, \log P \nonumber \\
                 &+8.56 (\pm0.80)\,(V-I)_0 \nonumber \\
                 &-0.40 (\pm0.15)
\end{align}

While the period term is very similar in both relations, the colour term differs in larger degree, which can be related to the colour shift
present in genuinely-contact systems which are in thermal equilibrium. However, the colour terms in the two equations are consistent with each
other, as well as with the one in the first relation for the entire population, within $2\sigma$ range. As for the period term this consistency is even 
within $1\sigma$ range. It is also worth noticing that the relation for near-contact systems shows tighter correlation with the $RMS$~=~0.40~mag,
whereas the one for genuinely-contact binaries is 0.55~mag. 

For a typical period and colour configuration for the studied sample (${\log}P = -0.1$, $(V-I)_0 = -0.2$) the $M_V$ estimations 
given by the two relations differs by 0.37~mag. This value, which is within $1\sigma$ range of both relations, tells how big
error is made if a wrong model is assumed.

\section{Discussion}

The relation obtained for early-type contact binaries from the LMC (Eq.~\eqref{eq:3}) differs significantly from the one known 
for late-type systems (Eq.~\eqref{eq:2}), even taking into account the relatively high uncertainty of the coefficients of
${\log}P$ and $(V-I)_0$ terms in the Eq.~\eqref{eq:3}. 
Fig.~8 presents the absolute magnitudes the original relation for late-type systems would yield if naively used for the sample of
early-type systems analysed here.
This clearly shows that the relation for the entire population is non-linear. The attempt to simply extend the results 
obtained for low-mass objects results in the $M_V$ estimation on average fainter by 1.2 mag.
The main reason of this is the bolometric correction, which is small or even negligible for late-type stars, becomes 
significant for, blue, hot stars. This results in the much steeper colour dependence for the 
early-type systems. However, it should be noted that the period dependence for massive stars also differs significantly from the one
for low-mass systems.

\begin{figure}
\label{fig:8}
\centerline{
\includegraphics[angle=270,width=86mm]{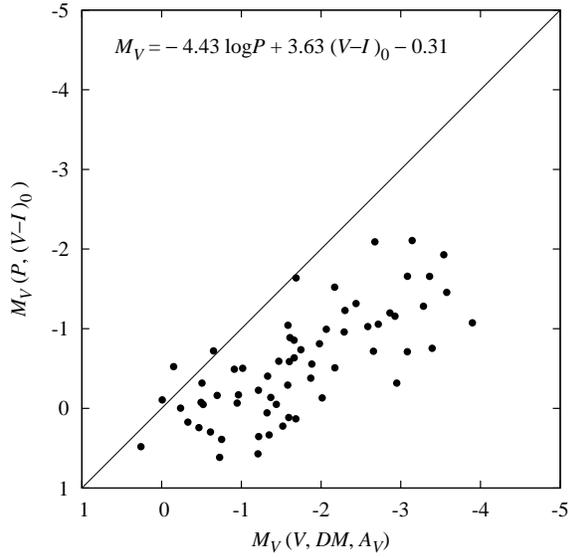}}
\caption{An attempt to naively extend the relation obtained by Rucinski (2004) for the late-type stars and use it for 
early-type objects. The absolute magnitudes calculated
using the distance to LMC are given on the horizontal axis, while the ones obtained from the PLC relation - on the vertical 
axis. It can be seen that relation obtained for late-type stars gives on average fainter absolute magnitudes than expected.}
\end{figure}

\begin{figure}
\label{fig:9}
\centerline{
\includegraphics[angle=270,width=86mm]{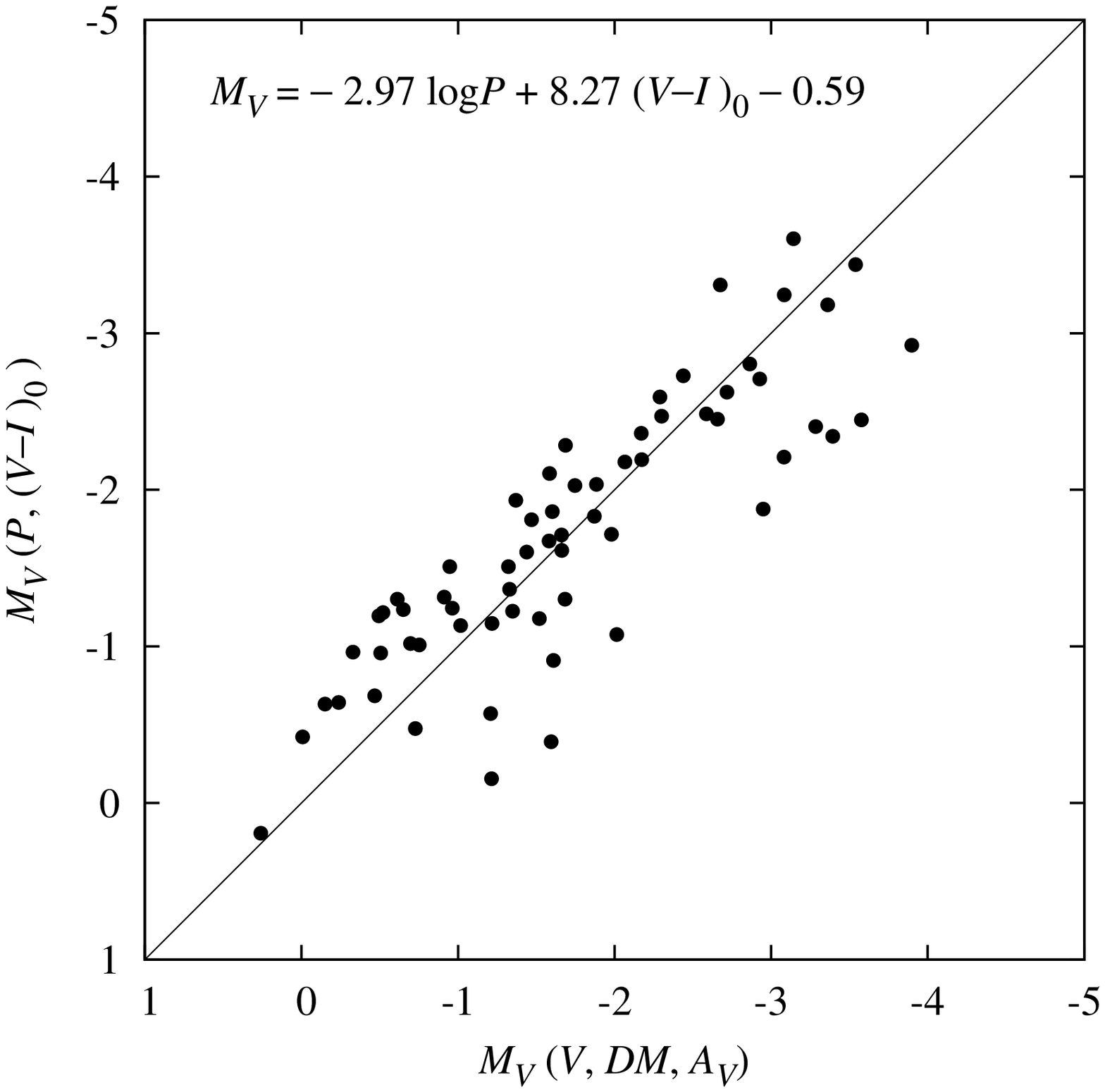}}
\centerline{
\includegraphics[angle=270,width=86mm]{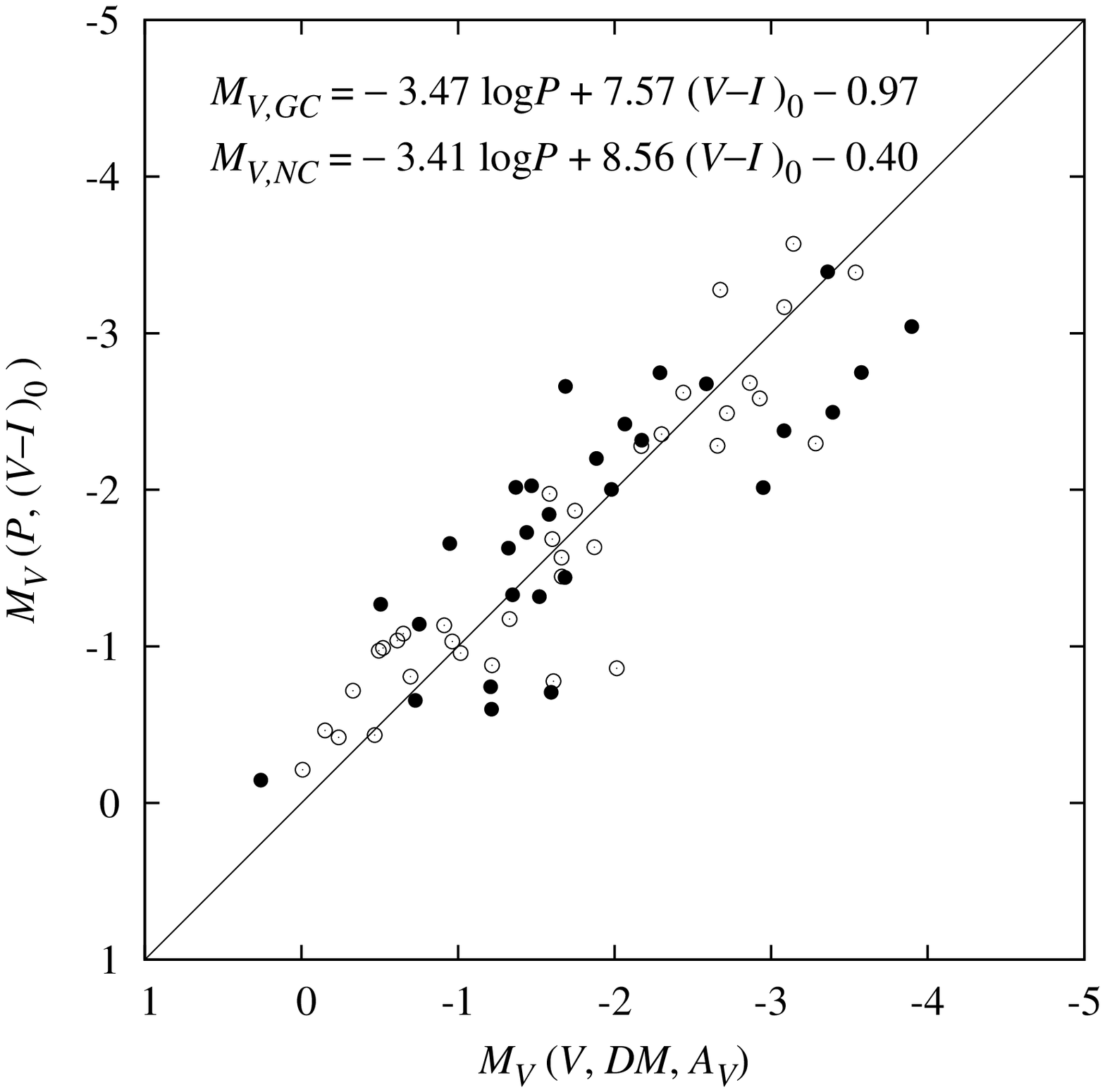}}
\caption{PLC relation fitted for the entire sample (upper panel) and genuinely-contact and near-contact systems separately (lower-panel). 
Description of the axes the same as in Fig.~8. The fit is slightly better when two separate relations are used.}
\end{figure}

While the formula obtained for the high-mass systems (Eq.~\eqref{eq:3}) fits the data (Fig.~9), the scatter of the points around 
it is large. The $RMS$ = 0.52 mag is large compared to the accuracy given by Rucinski (2004) for low-mass 
systems. This is related to a few factors. First, the obtained value of $M_V$ for systems in the LMC may not be as 
accurate as the values for nearby stars, especially due to the fact that reddening correction needs to be applied.
This introduces uncertainty of about 0.19 mag on average.

The assumption that all of the studied systems are at the same distance ($DM = 18.49$~mag) also contributes to the 
scatter of the relation. However, this contribution, while non-negligible, is much smaller than the uncertainty of
the reddening maps. For objects studied by Pietrzy{\'n}ski et al. (2013), where precise determination of distance to
individual objects was performed, the differences between the mean $DM$  to the LMC and to a given object 
are smaller than 0.04~mag. Even assuming the upper estimate
of the geometry-related uncertainty of 0.07~mag based on the scatter of the Classical Cepheids PL relation (Soszy{\'n}ski et al. 2008),
it is still much smaller than 0.19~mag, related to the reddening correction.

The large scatter of 0.52 mag relative to the predicted linear relation (Eq.~3) requires an explanation. Most likely, 
the main contributing factors are our lack of knowledge of the mass-ratio $q$ and our inability to distinguish 
between binaries in genuinely good contact from those which are lumped here under the name of near-contact ones. 
The latter may be versions of semi-detached binaries. For contact binaries, the efficient thermal transport equalizes 
the effective temperature over the common radiating surface. In contrast, the energy transport is apparently entirely 
or partially absent in near-contact binaries. The strong dependence on the configuration couples with the value of $q$. 
For genuine contact binaries, both the absolute magnitude and colour should be affected in a predictable way, while 
the combination of the magnitudes and colours is much harder to predict and not so obviously dependent on $q$ for near-contact binaries.

It should be noted that none of the objects used in this determination has a spectroscopically determined $q$ while photometrically 
determined values of $q$ are model dependent since light curve synthesis fits require assumption on the system configuration. 
For contact binaries, the dynamically permitted range extends from $q = 0.09$ (Rasio 1995, Arbutina 2007, 2009) to unity ($q=1$). 
For two identical stars, regardless of the model, the system should be simply two times brighter. Thus, the magnitude 
is then increased by 0.75, while the colour is unaffected. But, for any $q < 1$, the situation is different depending if 
the system is a genuinely-contact or a near-contact one. In a contact binary, for a decreasing value of $q$, the secondary 
component provides progressively less luminosity while it contributes a large amount of the common radiating area. In fact, 
the area stays relatively large down to small values of $q$. Thus, for smaller values of $q$, the result is a substantial 
modification of the colour of the contact binary with usually only a moderate modification of the absolute magnitude 
(Mochnacki (1981), Rucinski and Duerbeck (1997)). This picture applies only to binaries in perfect thermal contact. 
In contrast, we have currently no consistent model for near-contact binaries which appear to be common among early-type binaries.

Another factor ignored so far and that may potentially play a role here is metallicity. Rucinski (2002) suggests that a small
metallicity-dependent correction is necessary to obtain accurate absolute magnitude value. However, Rucinski (2004)
claims that explicit metallicity factor in the relation can be neglected, and PLC relation itself gives good
results for objects of different metallicity.

The objects in the studied sample belong to the same population and environment (the LMC bar and disc), therefore all of them
should have similar metallicity. Because of that the PLC relation obtained for them is expected to be self-consistent 
without including explicit metallicity factor. However, the difference in average metallicity of the LMC and 
Galactic stars may contribute to the difference of the relations obtained by Rucinski (2004) and in this work.

\section{Summary and Conclusions}

The analysis of early-type contact binaries from the LMC shows that they follow a PLC relation which is different 
from the one obtained for low-mass W UMa stars (Rucinski 2004). The relation obtained for the analysed systems 
is significantly steeper in the colour term than the one for late-type binaries and less steep in the period term. 
This leads to the conclusion that the relation for the entire population of contact binaries is non-linear.
While studied separately genuinely-contact and near-contact systems follow slightly different relations with 
the near-contact relation having smaller scatter.

While massive contact or close binaries clearly follow the PLC relation, its scatter is quite high. It is related to the 
physical properties of the system and a large diversity of possible configuration of system components and their mass
ratios. The main reason of the large $RMS$ is likely the lack of precise information on the $q$ value of a given system, which is 
impossible to obtain from the photometric data only. Different $q$ values lead to different 
magnitude and, in the case of genuinely-contact systems, colour shifts dependent on the type of
secondary component. For close, but non-contact systems where the second mentioned effect is absent, the scatter of the relation 
is smaller. Further study, including spectroscopic observations, may allow for more precise determination of the PLC relation.

There are also systematic uncertainties contributing to the scatter of the relation, especially related to the derredening
of the stars based on the extinction maps. While the metallicity factor can be neglected for the studied sample,
as it is expected to be similar for all studied objects, it may be necessary to take it into consideration while studying 
the whole population of W UMa stars.

\section*{Acknowledgements}
I would like to thank the referee, Prof. S{\l}awomir Ruci{\'n}ski for insightful comments and important contribution 
to this work.
I am also thankful to Prof. Igor Soszy{\'n}ski, Prof. Andrzej Udalski, Dr Jan Skowron, Dr Pawe{\l} Pietrukowicz,
and Dr Szymon Koz{\l}owski for fruitful discussion.
 
This work has been supported by Polish National Science Center grant No. 
DEC-2011/03/B/ST9/02573.

\clearpage

\begin{table}
\caption{Basic paramerers for the objects studied in this work. IDs are adopted from Graczyk et al. (2011).}
\label{tab:1}
\scalebox{0.9}{
\begin{tabular}[p!]{|l | c | c | c | c | c | l |}
\hline
ID & RA & DEC & $M_V$ & $(V-I)_0$ & log$P$ & type \\
\hline 
OGLE-LMC-ECL-00654 & 04:45:24.73 & -67:47:33.8 & -0.521 & -0.137 & -0.171301 & near-contact \\
OGLE-LMC-ECL-01278 & 04:49:53.60 & -68:58:50.3 & -1.661 & -0.129 &  0.017263 & near-contact \\
OGLE-LMC-ECL-01572 & 04:51:06.09 & -69:35:40.8 & -1.871 & -0.205 & -0.152281 & near-contact \\
OGLE-LMC-ECL-01725 & 04:51:39.44 & -70:42:22.5 & -1.584 & -0.175 &  0.022083 & near-contact \\
OGLE-LMC-ECL-01792 & 04:51:55.98 & -69:29:04.8 & -1.582 & -0.186 & -0.157848 & genuinely-contact \\
OGLE-LMC-ECL-02444 & 04:54:01.87 & -67:01:36.1 & -0.468 & -0.079 & -0.189630 & near-contact \\
OGLE-LMC-ECL-02776 & 04:54:52.22 & -67:08:03.4 & -0.728 & -0.087 & -0.279861 & genuinely-contact \\
OGLE-LMC-ECL-03026 & 04:55:32.53 & -69:28:27.1 & -3.083 & -0.300 &  0.058536 & near-contact \\
OGLE-LMC-ECL-03742 & 04:57:16.75 & -70:12:51.2 & -2.300 & -0.216 &  0.030166 & near-contact \\
OGLE-LMC-ECL-03778 & 04:57:20.43 & -67:02:24.3 & -2.950 & -0.220 & -0.178228 & genuinely-contact \\
OGLE-LMC-ECL-03780 & 04:57:20.58 & -68:52:37.8 & -3.284 & -0.199 &  0.056293 & near-contact \\
OGLE-LMC-ECL-03838 & 04:57:30.57 & -68:49:50.4 & -3.394 & -0.249 & -0.103829 & genuinely-contact \\
OGLE-LMC-ECL-04189 & 04:58:21.49 & -67:16:38.2 & -2.290 & -0.268 & -0.073589 & genuinely-contact \\
OGLE-LMC-ECL-04341 & 04:58:48.47 & -66:57:53.8 & -1.322 & -0.200 & -0.246135 & genuinely-contact \\
OGLE-LMC-ECL-05322 & 05:01:36.42 & -67:37:18.8 & -1.470 & -0.177 & -0.081093 & genuinely-contact \\
OGLE-LMC-ECL-05389 & 05:01:47.05 & -70:54:08.7 & -2.586 & -0.242 & -0.036790 & genuinely-contact \\
OGLE-LMC-ECL-05906 & 05:03:00.41 & -70:20:38.8 & -2.927 & -0.265 & -0.026323 & near-contact \\
OGLE-LMC-ECL-05977 & 05:03:10.27 & -67:52:00.5 & -0.151 &  0.018 &  0.062884 & near-contact \\
OGLE-LMC-ECL-06062 & 05:03:20.89 & -68:58:26.4 & -0.008 &  0.005 & -0.041275 & near-contact \\
OGLE-LMC-ECL-06121 & 05:03:30.51 & -69:54:54.7 & -0.752 & -0.152 & -0.282502 & genuinely-contact \\
OGLE-LMC-ECL-06667 & 05:04:43.35 & -70:19:48.3 & -1.349 & -0.183 & -0.294782 & near-contact \\
OGLE-LMC-ECL-06731 & 05:04:50.84 & -70:08:58.9 & -3.361 & -0.289 &  0.067021 & genuinely-contact \\
OGLE-LMC-ECL-06745 & 05:04:52.91 & -70:33:01.8 & -0.948 & -0.186 & -0.207001 & genuinely-contact \\
OGLE-LMC-ECL-07009 & 05:05:29.38 & -70:31:53.4 & -2.718 & -0.263 & -0.046756 & near-contact \\
OGLE-LMC-ECL-07050 & 05:05:35.36 & -67:03:11.6 &  0.260 &  0.043 & -0.142969 & genuinely-contact \\
OGLE-LMC-ECL-07378 & 05:06:14.93 & -69:09:02.9 & -0.507 & -0.062 & -0.049429 & genuinely-contact \\
OGLE-LMC-ECL-08338 & 05:08:32.69 & -71:13:38.5 & -3.577 & -0.186 &  0.106351 & genuinely-contact \\
OGLE-LMC-ECL-08909 & 05:09:51.75 & -69:02:03.9 & -2.171 & -0.164 &  0.138865 & near-contact \\
OGLE-LMC-ECL-09001 & 05:10:04.09 & -69:37:02.2 & -0.238 & -0.044 & -0.105518 & near-contact \\
OGLE-LMC-ECL-09055 & 05:10:11.76 & -69:06:25.6 & -1.208 & -0.098 & -0.279223 & genuinely-contact \\
OGLE-LMC-ECL-09283 & 05:10:46.07 & -67:36:11.5 & -2.864 & -0.277 & -0.026775 & near-contact \\
OGLE-LMC-ECL-09482 & 05:11:14.51 & -69:16:18.1 & -1.980 & -0.135 &  0.002582 & genuinely-contact \\
OGLE-LMC-ECL-09689 & 05:11:52.92 & -69:11:25.6 & -1.884 & -0.219 & -0.123840 & genuinely-contact \\
OGLE-LMC-ECL-09787 & 05:12:06.83 & -68:56:45.0 & -0.329 & -0.119 & -0.206667 & near-contact \\
OGLE-LMC-ECL-10351 & 05:13:32.97 & -68:49:57.4 & -0.697 & -0.090 & -0.107291 & near-contact \\
OGLE-LMC-ECL-10542 & 05:13:59.88 & -69:04:55.7 & -2.675 & -0.261 &  0.188063 & near-contact \\
OGLE-LMC-ECL-11003 & 05:15:07.52 & -66:38:24.7 & -1.747 & -0.197 & -0.065596 & near-contact \\
OGLE-LMC-ECL-11541 & 05:16:28.73 & -67:53:28.9 & -3.539 & -0.302 &  0.117821 & near-contact \\
OGLE-LMC-ECL-11769 & 05:17:03.35 & -69:42:04.8 & -2.065 & -0.194 & -0.004378 & genuinely-contact \\
OGLE-LMC-ECL-12030 & 05:17:43.90 & -67:55:20.4 & -2.173 & -0.252 & -0.161479 & genuinely-contact \\
OGLE-LMC-ECL-12415 & 05:18:35.41 & -67:51:23.9 & -3.142 & -0.310 &  0.151689 & near-contact \\
OGLE-LMC-ECL-12649 & 05:19:10.04 & -68:00:37.3 & -1.369 & -0.250 & -0.243720 & genuinely-contact \\
OGLE-LMC-ECL-12802 & 05:19:33.33 & -69:04:37.1 & -1.610 &  0.012 &  0.140016 & near-contact \\
OGLE-LMC-ECL-12866 & 05:19:44.67 & -67:53:30.5 & -1.595 & -0.015 & -0.108202 & genuinely-contact \\
OGLE-LMC-ECL-12878 & 05:19:45.70 & -71:14:19.2 & -3.082 & -0.231 & -0.099056 & genuinely-contact \\
OGLE-LMC-ECL-12982 & 05:19:59.86 & -68:15:09.2 & -3.898 & -0.312 & -0.083027 & genuinely-contact \\
OGLE-LMC-ECL-13206 & 05:20:36.00 & -69:26:09.1 & -1.602 & -0.186 & -0.089552 & near-contact \\
OGLE-LMC-ECL-14291 & 05:23:05.58 & -69:35:03.1 & -1.329 & -0.122 & -0.078460 & near-contact \\
OGLE-LMC-ECL-14617 & 05:23:49.28 & -69:14:21.9 & -2.439 & -0.251 &  0.021698 & near-contact \\
OGLE-LMC-ECL-15444 & 05:25:54.31 & -69:27:24.1 & -1.684 & -0.172 & -0.240697 & genuinely-contact \\
OGLE-LMC-ECL-15756 & 05:26:35.01 & -69:52:14.6 & -0.495 & -0.131 & -0.160075 & near-contact \\
OGLE-LMC-ECL-16083 & 05:27:11.41 & -69:50:53.3 & -1.686 & -0.138 &  0.187106 & genuinely-contact \\
OGLE-LMC-ECL-18264 & 05:31:36.39 & -68:59:24.7 & -1.520 & -0.162 & -0.252703 & genuinely-contact \\
OGLE-LMC-ECL-18865 & 05:32:56.86 & -70:44:37.2 & -1.438 & -0.203 & -0.225302 & genuinely-contact \\
OGLE-LMC-ECL-19039 & 05:33:19.72 & -69:37:19.7 & -1.663 & -0.138 & -0.039738 & near-contact \\
OGLE-LMC-ECL-20231 & 05:36:02.21 & -69:24:32.3 & -2.657 & -0.272 & -0.130987 & near-contact \\
OGLE-LMC-ECL-21029 & 05:37:46.54 & -67:47:36.9 & -0.912 & -0.103 & -0.043745 & near-contact \\
OGLE-LMC-ECL-21066 & 05:37:50.51 & -71:39:53.7 & -1.218 & -0.172 & -0.290562 & near-contact \\
OGLE-LMC-ECL-21744 & 05:39:32.41 & -71:32:21.8 & -1.017 & -0.071 & -0.014624 & near-contact \\
OGLE-LMC-ECL-21983 & 05:40:05.29 & -68:28:35.2 & -1.214 &  0.065 &  0.034789 & genuinely-contact \\
OGLE-LMC-ECL-22275 & 05:40:47.88 & -70:01:27.9 & -2.014 & -0.104 & -0.125227 & near-contact \\
OGLE-LMC-ECL-23809 & 05:45:24.37 & -69:38:21.7 & -0.613 & -0.191 & -0.293656 & near-contact \\
OGLE-LMC-ECL-24776 & 05:50:00.45 & -70:09:06.2 & -0.964 & -0.128 & -0.136344 & near-contact \\
OGLE-LMC-ECL-25114 & 05:52:18.19 & -71:53:42.9 & -0.651 & -0.063 &  0.040922 & near-contact \\
\hline
\end{tabular}}
\end{table}

\end{document}